\pdfoutput=1

\documentclass[british, pra, twocolumn]{revtex4-1}
\usepackage[T1]{fontenc}
\usepackage[latin9]{inputenc}
\usepackage{color}
\usepackage{babel}
\usepackage{bm}
\usepackage{amsmath}
\usepackage{amssymb}
\usepackage{graphicx}
\usepackage[unicode=true,pdfusetitle,
 bookmarks=true,bookmarksnumbered=false,bookmarksopen=false,
 breaklinks=false,pdfborder={0 0 0},pdfborderstyle={},backref=false,colorlinks=true]
 {hyperref}
\hypersetup{
 linkcolor = blue, citecolor = blue, urlcolor = blue}
\begin{document}

\title{Electronic spectral properties of incommensurate twisted trilayer
graphene}

\author{B. Amorim$^{1}$, Eduardo V. Castro$^{1,2}$}

\affiliation{$^{1}$CeFEMA, Instituto Superior Técnico, Universidade de Lisboa,
Av. Rovisco Pais, 1049-001 Lisboa, Portugal}

\affiliation{$^{2}$Centro de Física das Universidades do Minho e Porto, Departamento
de Física e Astronomia, Faculdade de Ciências, Universidade do Porto,
4169-007 Porto, Portugal}
\begin{abstract}
Multilayered van der Waals structures often lack periodicity, which
difficults their modeling. Building on previous work for bilayers,
we develop a tight-binding based, momentum space formalism capable
of describing incommensurate multilayered van der Waals structures
for arbitrary lattice mismatch and/or misalignment between different
layers. We demonstrate how the developed formalism can be used to
model angle-resolved photoemission spectroscopy measurements, and
scanning tunnelling spectroscopy which can probe the local and total
density of states. The general method is then applied to incommensurate
twisted trilayer graphene structures. It is found that the coupling
between the three layers can significantly affect the low energy spectral
properties, which cannot be simply attributed to the pairwise hybridization
between the layers.
\end{abstract}
\maketitle

\section{Introduction}

The rise of two-dimensional (2D) materials, in recent years, has enabled
the study of structures formed by vertically stacked 2D layers\cite{Ponomarenko_2011,Novoselov_2012,Geim_2013}.
This new kind of structures, generally referred to as van der Waals
(vdW) structures due to the interaction that holds the layers together,
display new and interesting physics. The properties of vdW structures
are determined not only by the properties of the individual layers,
but also, sometimes in a fundamental way, by the coupling between
different layers, which is affected by the relative lattice mismatch
and misalignment. 

A prototypical van der Waals structure is twisted bilayer graphene
(tBLG). This apparently simple material displays rich and interesting
properties that deviate substantially from both single layer and Bernal
stacked bilayer graphene. The misalignment between the two layers,
which gives origin to moiré patterns, is responsible for the reduction
of graphene's Fermi velocity\cite{LopesDosSantos_2007,Yufeng_2010,Luican_2011,Trambly_2010}
and to the emergence of low energy van Hove singularities\cite{LopesDosSantos_2007,Li_2009},
both of which are controlled by the twist angle. For very small twist
angles, the van Hove singularities that occur above and bellow graphene's
neutrality point can coalesce, leading to the formation of the so
called flat bands at the neutrality point\cite{Trambly_2010,Morell_2010,Bistritzer_2011}.
Very recently, a strongly correlated Mott insulating phase\cite{Cao_2018a}
and superconductivity\cite{Cao_2018b} have been observed in tBLG
in the flat band regime. The effect of twist has also been observed
in semiconducting transition metal dichalcogenides (STMD). Namelly,
it was found that the band gap, and whether it is direct or indirect,
of bilayer MoS$_{2}$ is controlled by the relative twist angle\cite{Yeh_2016}. 

The study of van der Waals structures is not only of fundamental interested,
but also has potential technological applications. Hybrid vertical
structures formed by graphene/boron nitride/graphene have been shown
to display negative differential conductance in their vertical transport
characteristics, which can be exploited to create a radio-frequency
oscillator\cite{Mishchenko_2014}. Graphene/boron nitride/graphene\cite{Britnell_2012,Britnell_2012b}
and graphene/STMD/graphene structures\cite{Britnell_2012,Georgiou_2012}
were also shown to operate as vertical tunneling field effect transistors
with large ON/OFF ratios. Graphene/STMD/graphene structures can also
be used as photodetectors with fast response times\cite{Britnell_2013,Yu_2013,Massicotte_2015}. 

The possible lattice mismatch/misalignment in van der Waals structures
and the frequent sensitivity of their properties to those, makes the
modelling of such structures challenging. The lattice mismatch/misalignment
can give origin to periodic structures with large unit cells, making
treatments based on Bloch's theorem numerically expensive. In the
case when the structure is incommensurate, Bloch's theorem cannot
be applied. For the case of tBLG, a momentum space formalism, based
on the expansion of the electronic wave function in Bloch states of
the individual layers, which can undergo generalized umklapp scattering,
has been developed\cite{LopesDosSantos_2007,Shallcross_2008,Bistritzer_2010,Bistritzer_2011,LopesDosSantos_2012,Moon_2013,Koshino_2015}
(a mathematically formal description of the method can be found in
\cite{Massatt_2018}). This method has proved to be very useful, allowing
to model incommensurate or commensurate, large period structures,
at a modest computational cost. This method is not restricted to tBLG,
but can be applied to other kinds structures even for large mismatch/misaligment
\cite{Koshino_2015,Amorim_2018}. 

Theoretical work up to now has been focused on the study of incommensurate
bilayer structures (formed by two lattice mismatched periodic structures).
An exception to this is Ref.~\cite{Correa_2014}, where the optical
properties of commensurate fully twisted trilayer graphene (tTLG),
where all layers are rotated, are studied using \textit{ab initio}
methods. However, the twisted trilayer structures that can be easily
simulated is even more restricted than in the bilayer case, due to
the even larger unit cells involved. The interest in lattice mismatched/misaligned
multilayer structures, such as graphene/boron nitride/graphene and
graphene/STMC/graphene, demands the development of numerically efficient
methods. The goal of this paper is to extend the momentum space method
to multilayer incommensurate structures. We study how the electronic
wavefunctions, and the corresponding energies, can be determined and
how these can be used to evaluate different measurable spectral quantities,
namely, the angle-resolved photoemission spectroscopy (ARPES) intensity,
and the total (TDoS) and local densities of states (LDoS), which can
be measured via scanning tunnelling spectroscopy (STS). Motivated
by recent experimental work \cite{Zuo_2018}, we use the developed
method to study the spectral properties of tTLG. 

This paper is organized as follows. The general formalism in developed
in Section~\ref{sec:Formalism}. In Section~\ref{subsec:Hamiltonian},
an effective Hamiltonian in momentum space for incommensurate multilayers
is constructed. In Section~\ref{subsec:Spectral}, it is exemplified
how the effective Hamiltonian can be used to model ARPES measurements,
LDoS and TDoS. The general formalism is applied to tTLG in Section
\ref{sec:Application_tTLG}. Finally, we conclude in Section~\ref{sec:Conclusions}
and also discuss future uses of the formalism.

\section{Formalism\label{sec:Formalism}}

\subsection{Hamiltonian for incommensurate multilayers in momentum space\label{subsec:Hamiltonian}}

For simplicity, we will specialize to the case of trilayer structures,
which already captures all the conceptual complexity of a multilayer.
To clarify, by a trilayer we mean a structure that is formed by three
periodic systems which are lattice mismatched/misaligned. In this
way, the structure formed by a Bernall-staked graphene bilayer with
an additional twisted graphene monolayer, as studied in Ref.~\cite{Morell_2013},
would be classified as a bilayer. In order to model the electronic
properties of incommensurate, lattice mismatched/misaligned multilayer
structures, we start from a tight-binding description of the system,
as previously done for bilayers\cite{LopesDosSantos_2007,Bistritzer_2010,Koshino_2015}.
The single-electron Hamiltonian of the trilayer reads
\begin{equation}
H=\sum_{\ell=1}^{3}H_{\ell}+\sum_{\ell=1}^{2}\left(H_{\ell+1,\ell}+H_{\ell,\ell+1}\right),\label{eq:Hamiltonian}
\end{equation}
where $H_{\ell}$ ($\ell=1,...,3$) describe the isolated layers,
which are assumed to be periodic, and $H_{\ell,\ell^{\prime}}$ describe
the hopping of electrons from layer $\ell^{\prime}$ to layer $\ell$,
with $H_{\ell^{\prime},\ell}=H_{\ell,\ell^{\prime}}^{\dagger}$. Owing
to the exponential suppression of the hopping integrals with distance,
we have assumed that only consecutive layers are coupled to each other.
In terms of creation and annihilation operators, we write the intralayer
Hamiltonians as
\begin{equation}
H_{\ell}=\sum_{\mathbf{R}_{\ell}\alpha,\mathbf{R}_{\ell}^{\prime}\alpha^{\prime}}h_{\alpha\alpha^{\prime}}^{\ell\ell}\left(\mathbf{R}_{\ell},\mathbf{R}_{\ell}^{\prime}\right)c_{\ell,\mathbf{R}_{\ell},\alpha}^{\dagger}c_{\ell,\mathbf{R}_{\ell}^{\prime},\alpha^{\prime}},
\end{equation}
where $h_{\alpha\alpha^{\prime}}^{\ell\ell}\left(\mathbf{R}_{\ell},\mathbf{R}_{\ell}^{\prime}\right)=h_{\alpha\alpha^{\prime}}^{\ell\ell}\left(\mathbf{R}_{\ell}-\mathbf{R}_{\ell}^{\prime},\mathbf{0}\right)$
are hopping parameters, which are invariant under lattice translations
of layer $\ell$, and $c_{\ell,\mathbf{R}_{\ell},\alpha}^{\dagger}$
creates an electron in a Wannier state of orbital/sublattice character
$\alpha$, which is centered at the position $\mathbf{R}_{\ell}+\bm{\tau}_{\ell,\alpha}$,
with $\mathbf{R}_{\ell}$ a Bravais lattice site of layer $\ell$
and $\bm{\tau}_{\ell,\alpha}$ the position of the Wannier center
within the unit cell. We represent the Wannier states as $\left|\ell,\mathbf{R}_{\ell},\alpha\right\rangle $
and $N_{\text{orb}\ell}$ is the number of Wannier orbitals per unit
cell of layer $\ell$. The interlayer terms of the Hamiltonian read
\begin{equation}
H_{\ell,\ell^{\prime}}=\sum_{\mathbf{R}_{\ell}\alpha,\mathbf{R}_{\ell}^{\prime}\beta}h_{\alpha\beta}^{\ell\ell^{\prime}}\left(\mathbf{R}_{\ell},\mathbf{R}_{\ell^{\prime}}\right)c_{\ell,\mathbf{R}_{\ell},\alpha}^{\dagger}c_{\ell^{\prime},\mathbf{R}_{\ell^{\prime}},\beta},
\end{equation}
where $h_{\alpha\beta}^{\ell\ell^{\prime}}\left(\mathbf{R}_{\ell},\mathbf{R}_{\ell^{\prime}}\right)$
are interlayer hopping terms, with $\mathbf{R}_{\ell}$($\mathbf{R}_{\ell^{\prime}}$)
running over lattice sites of layer $\ell$($\ell^{\prime}$) and
$\alpha$($\beta$) running over the orbital/sublattice degrees of
freedom of layer $\ell$($\ell^{\prime}$). The Wannier states can
be written of Bloch waves, $\left|\ell,\mathbf{k},\alpha\right\rangle $,
of the individual layers as
\begin{equation}
\left|\ell,\mathbf{R}_{\ell},\alpha\right\rangle =\frac{1}{\sqrt{N_{\ell}}}\sum_{\mathbf{k}\in\text{BZ}\ell}e^{-i\mathbf{k}\cdot\left(\mathbf{R}_{\ell}+\bm{\tau}_{\ell,\alpha}\right)}\left|\ell,\mathbf{k},\alpha\right\rangle ,
\end{equation}
where $\text{BZ}\ell$ represents the Brillouin zone for layer $\ell$
and $N_{\ell}$ is the number of unit cells in layer $\ell$. Changing
to the Bloch wave basis brings the Hamiltonians of the isolated layers
to a block diagonal form
\begin{equation}
H_{\ell}=\sum_{\mathbf{k}\in\text{BZ}\ell,\alpha\alpha^{\prime}}h_{\alpha\alpha^{\prime}}^{\ell\ell}\left(\mathbf{k}\right)c_{\ell,\mathbf{k},\alpha}^{\dagger}c_{\ell,\mathbf{k},\alpha^{\prime}},
\end{equation}
where $h_{\alpha\alpha^{\prime}}^{\ell\ell}\left(\mathbf{k}\right)=\sum_{\mathbf{R}_{\ell}}e^{-i\mathbf{k}\cdot\left(\mathbf{R}_{\ell}+\bm{\tau}_{\ell,\alpha}-\bm{\tau}_{\ell,\beta}\right)}h_{\alpha\beta}^{\ell\ell}\left(\mathbf{R}_{\ell},\mathbf{0}\right)$
and $c_{\ell,\mathbf{k},\alpha}^{\dagger}$ creates an electron in
the Bloch state $\left|\ell,\mathbf{k},\alpha\right\rangle $. Assuming
a two-centre approximation for the interlayer hoppings, these can
be written as a Fourier transform \cite{Bistritzer_2010,Koshino_2015}
\begin{multline}
h_{\alpha\beta}^{\ell\ell^{\prime}}\left(\mathbf{R}_{\ell},\mathbf{R}_{\ell^{\prime}}\right)=\sqrt{A_{\text{u.c.}\ell}A_{\text{u.c.}\ell^{\prime}}}\times\\
\times\int\frac{d^{2}\mathbf{q}}{\left(2\pi\right)^{2}}e^{i\mathbf{q}\cdot\left(\mathbf{R}_{\ell}+\bm{\tau}_{\ell,\alpha}-\mathbf{R}_{\ell^{\prime}}-\bm{\tau}_{\ell^{\prime},\beta}\right)}h_{\alpha\beta}^{\ell\ell^{\prime}}\left(\mathbf{q}\right),
\end{multline}
and the interlayer terms of the Hamiltonian can be written as
\begin{multline}
H_{\ell,\ell^{\prime}}=\sum_{\substack{\mathbf{k}\in\text{BZ}\ell,\alpha,\mathbf{G}_{\ell}\\
\mathbf{k}^{\prime}\in\text{BZ}\ell^{\prime},\beta,\mathbf{G}_{\ell^{\prime}}
}
}e^{i\mathbf{G}_{\ell}\cdot\bm{\tau}_{\ell,\alpha}}h_{\alpha\beta}^{\ell\ell^{\prime}}\left(\mathbf{k}+\mathbf{G}_{\ell}\right)e^{-i\mathbf{G}_{\ell^{\prime}}\cdot\bm{\tau}_{\ell^{\prime},\beta}}\\
\times c_{\ell,\mathbf{k},\alpha}^{\dagger}c_{\ell^{\prime},\mathbf{k}^{\prime},\beta}\delta_{\mathbf{k}+\mathbf{G}_{\ell},\mathbf{k}^{\prime}+\mathbf{G}_{\ell^{\prime}}},
\end{multline}
where $\mathbf{G}_{\ell}$($\mathbf{G}_{\ell^{\prime}}$) are reciprocal
lattice vectors of layer $\ell$($\ell^{\prime}$). The Kronecher
symbol in the above equation imposes the generalized umklapp condition
\cite{Koshino_2015}, which states that two Bloch states of layer
$\ell$ and $\ell^{\prime}$ with cystal-momentum $\mathbf{k}$ and
$\mathbf{k}^{\prime}$, respectively, are only coupled to each other
provided reciprocal lattice vectors of layers $\ell$, $\mathbf{G}_{\ell}$
, and $\ell^{\prime}$, $\mathbf{G}_{\ell^{\prime}}$, exist such
that $\mathbf{k}+\mathbf{G}_{\ell}=\mathbf{k}^{\prime}+\mathbf{G}_{\ell^{\prime}}$.
This condition must be satisfied for each hopping process between
two consecutive layers. 

Now let us study when two Bloch states of non-consecutive layers,
in a multilayer structure, can couple by compounding generalized umklapp
processes. Let us consider a trilayer structure. We have a state with
momentum $\mathbf{k}_{1}$ of layer $1$, which couples to a state
of layer $2$ with momentum $\mathbf{k}_{2}$, provided reciprocal
vectors $\mathbf{G}_{1}$ and $\mathbf{G}_{2}$ exist such that $\mathbf{k}_{1}+\mathbf{G}_{1}=\mathbf{k}_{2}+\mathbf{G}_{2}.$
In its turn, state $\mathbf{k}_{2}$ can couple to a state of layer
$3$ with momentum $\mathbf{k}_{3}$, provided $\mathbf{G}_{2}^{\prime}$
and $\mathbf{G}_{3}$ exist, such that $\mathbf{k}_{2}+\mathbf{G}_{2}^{\prime}=\mathbf{k}_{3}+\mathbf{G}_{3}.$
Therefore, states $\mathbf{k}_{1}$ and $\mathbf{k}_{3}$ are coupled,
provided reciprocal lattice vectors $\mathbf{G}_{1}$ of layer $1$,
$\mathbf{G}_{3}$ of layer $3$, $\mathbf{G}_{2}$ and $\mathbf{G}_{2}^{\prime}$
of layer $2$ exist, such that
\begin{equation}
\mathbf{k}_{1}+\mathbf{G}_{1}+\mathbf{G}_{2}^{\prime}=\mathbf{k}_{3}+\mathbf{G}_{3}+\mathbf{G}_{2},
\end{equation}
where $\mathbf{G}_{2}$ and $\mathbf{G}_{2}^{\prime}$ can differ.
This is immediately satisfied, by working in a extended zone scheme
and setting $\mathbf{k}_{1}=\mathbf{p}+\mathbf{G}_{2}+\mathbf{G}_{3}$
and $\mathbf{k}_{3}=\mathbf{p}+\mathbf{G}_{1}+\mathbf{G}_{2}^{\prime}$,
with $\mathbf{p}$ defined in the extended reciprocal space. This
motivates us to look for eigenstates of Eq.~(\ref{eq:Hamiltonian})
of the form
\begin{multline}
\left|\psi_{\mathbf{k},n}^{\text{umklapp}}\right\rangle =\\
=\sum_{\mathbf{G}_{2},\mathbf{G}_{3},\alpha}\phi_{1,\mathbf{k},\alpha}^{n}\left(\mathbf{G}_{2},\mathbf{G}_{3}\right)\left|1,\mathbf{k}+\mathbf{G}_{2}+\mathbf{G}_{3},\alpha\right\rangle \\
+\sum_{\mathbf{G}_{1},\mathbf{G}_{3},\beta}\phi_{2,\mathbf{k},\beta}^{n}\left(\mathbf{G}_{1},\mathbf{G}_{3}\right)\left|2,\mathbf{k}+\mathbf{G}_{1}+\mathbf{G}_{3},\beta\right\rangle \\
+\sum_{\mathbf{G}_{1},\mathbf{G}_{2},\gamma}\phi_{3,\mathbf{k},\gamma}^{n}\left(\mathbf{G}_{1},\mathbf{G}_{2}\right)\left|3,\mathbf{k}+\mathbf{G}_{1}+\mathbf{G}_{2},\gamma\right\rangle ,\label{eq:umklapp_eigenstate}
\end{multline}
which is a superposition of Bloch states of the three layers, where
the Bloch state from one layer can undergo generalized umklapp scattering
due to the remaining two layers. The generalization for the multilayer
case is formally straightforward: the multilayer eigenstates are formed
by a superposition of Bloch states of each layer, which can undergo
umklapp scattering by reciprocal lattice vectors of all the remaining
ones. 

By suitably truncating the sums over reciprocal lattice vectors in
Eq.~(\ref{eq:umklapp_eigenstate}), we obtain an effective Hamiltonian
which can be written as
\begin{equation}
\bm{H}_{\mathbf{k}}^{\text{umklapp}}=\left[\begin{array}{ccc}
\bm{H}_{\mathbf{k}}^{11} & \bm{H}_{\mathbf{k}}^{12} & \bm{0}\\
\bm{H}_{\mathbf{k}}^{21} & \bm{H}_{\mathbf{k}}^{22} & \bm{H}_{\mathbf{k}}^{23}\\
\bm{0} & \bm{H}_{\mathbf{k}}^{32} & \bm{H}_{\mathbf{k}}^{33}
\end{array}\right],\label{eq:umklapp_hamiltonian}
\end{equation}
with the matrix entries running over $\mathbf{G}_{2},\mathbf{G}_{3},\alpha$
for the layer $1$ sector, and equivalently for the two other layers.
In the above expression, $\bm{H}_{\mathbf{k}}^{11}$ is a block diagonal
matrix, with entries given by
\begin{multline}
\left[\bm{H}_{\mathbf{k}}^{11}\right]_{\mathbf{G}_{2},\mathbf{G}_{3},\alpha;\mathbf{G}_{2}^{\prime},\mathbf{G}_{3}^{\prime},\beta}=\delta_{\mathbf{G}_{2},\mathbf{G}_{2}^{\prime}}\delta_{\mathbf{G}_{3},\mathbf{G}_{3}^{\prime}}\\
\times h_{\alpha\beta}^{11}\left(\mathbf{k}+\mathbf{G}_{2}+\mathbf{G}_{3}\right),
\end{multline}
and similarly for $\bm{H}_{\mathbf{k}}^{22}$ and $\bm{H}_{\mathbf{k}}^{33}$.
For the interlayer terms we have
\begin{multline}
\left[\bm{H}_{\mathbf{k}}^{12}\right]_{\mathbf{G}_{2},\mathbf{G}_{3},\alpha;\mathbf{G}_{1}^{\prime},\mathbf{G}_{3}^{\prime},\beta}=\delta_{\mathbf{G}_{3},\mathbf{G}_{3}^{\prime}}\\
\times e^{i\mathbf{G}_{1}^{\prime}\cdot\bm{\tau}_{1,\alpha}}h_{\alpha\beta}^{12}\left(\mathbf{k}+\mathbf{G}_{3}+\mathbf{G}_{2}+\mathbf{G}_{1}^{\prime}\right)e^{-i\mathbf{G}_{2}\cdot\bm{\tau}_{2,\alpha}},
\end{multline}
where the $\delta_{\mathbf{G}_{3},\mathbf{G}_{3}^{\prime}}$ emerges
due to the fact that in a hopping process between layers $1$ and
$2$, only exchanges of momentum by reciprocal lattice vectors of
layers 1 and 2 are involved and we are assuming that the structure
is incommensurate. $\bm{H}_{\mathbf{k}}^{23}$ is constructed in a
similar way and $\bm{H}_{\mathbf{k}}^{\ell^{\prime}\ell}=\left[\bm{H}_{\mathbf{k}}^{\ell\ell^{\prime}}\right]^{\dagger}$.
In order to construct a finite matrix $\bm{H}_{\mathbf{k}}^{\text{umklapp}}$
it is necessary to impose a criterion to truncate the number of reciprocal
vectors involved. We notice that (i) the functions $h_{\alpha\beta}^{\ell\ell^{\prime}}\left(\mathbf{q}\right)$
decay very fast for large values of $\left|\mathbf{q}\right|$, (ii)
as we will see in the following, several observables are dominated
by the coefficients $\phi_{\ell,\mathbf{k},\alpha}^{n}\left(\mathbf{0},\mathbf{0}\right)$,
and (iii) in perturbation theory a scattering process $\left|1,\mathbf{k},\alpha\right\rangle \rightarrow\left|1,\mathbf{k}+\mathbf{G}_{2},\alpha\right\rangle $
is of second order in the interlayer coupling, while a scattering
process $\left|1,\mathbf{k},\alpha\right\rangle \rightarrow\left|1,\mathbf{k}+\mathbf{G}_{2}+\mathbf{G}_{3},\alpha\right\rangle $
is of fourth order. These three facts motivate us to only include
coefficients $\phi_{1,\mathbf{k},\alpha}^{n}\left(\mathbf{G}_{2},\mathbf{G}_{3}\right)$
such that $\left|\mathbf{G}_{2}\right|,\left|\mathbf{G}_{3}\right|,\left|\mathbf{G}_{2}+\mathbf{G}_{3}\right|<\Lambda$,
where $\Lambda$ is a momentum cutoff that controls the accuracy of
the calculation. The same criterion is applied to the coefficients
$\phi_{2,\mathbf{k},\alpha}^{n}\left(\mathbf{G}_{1},\mathbf{G}_{3}\right)$
and $\phi_{3,\mathbf{k},\alpha}^{n}\left(\mathbf{G}_{1},\mathbf{G}_{2}\right)$.
Diagonalizing the Hamiltonian Eq.~(\ref{eq:umklapp_hamiltonian}),
we obtain the eigenstates $\left|\psi_{\mathbf{k},n}^{\text{umklapp}}\right\rangle $
and the corresponding energies $E_{\mathbf{k},n}$, which can be used
to evaluate different physical observables. 

\subsection{Spectral observables\label{subsec:Spectral}}

We will now determine how the formalism described in the previous
section can be used to evaluate spectral quantities of incommensurate
multilayer systems. These quantities can be obtained by projecting
the spectral function $A(\omega)=\delta\left(\omega-H\right)$ against
suitable states. 

\subsubsection{Angle-resolved photoemission}

Angle-resolved photoemission spectroscopy allows to probe the momentum
resolved density of states of the system. In a periodic system, it
provides information about the electronic band structure. We will
extend the approach of Ref.~\cite{Amorim_2018} to model ARPES in
incommensurate bilayer structures to the multilayer case. However,
the approach followed here will be slightly different. The starting
point is the Fermi's golden rule-like expression for the energy resolved
ARPES intensity of photoemitted electrons with energy $E$ and momentum
$\mathbf{p}$, given that the electronic system was illuminated with
radiation with frequency $\omega_{0}$ and wavevector $\mathbf{q}$
\cite{Caroli_1973,Feibelman_1974,Amorim_2018}. To second order in
the radiation field, we have
\begin{multline}
I_{\text{ARPES}}\left(E,\mathbf{p}|\omega_{0},\mathbf{q}\right)\propto f\left(\omega-\mu\right)\\
\times\sum_{\substack{\ell\mathbf{R}_{\ell}\alpha\\
\ell^{\prime}\mathbf{R}_{\ell^{\prime}}\alpha^{\prime}
}
}M_{E,\mathbf{p}\mid\omega_{0},\mathbf{q}\mid\ell\mathbf{R}_{\ell}\alpha}A_{\ell\mathbf{R}_{\ell}\alpha;\ell^{\prime}\mathbf{R}_{\ell^{\prime}}\alpha^{\prime}}(\omega)M_{E,\mathbf{p}\mid\omega_{0},\mathbf{q}\mid\ell^{\prime}\mathbf{R}_{\ell^{\prime}}\alpha^{\prime}}^{*},\label{eq:ARPEs_fermi_golden_rule}
\end{multline}
where $\omega=E-\omega_{0}$, $f(\omega)=\left(e^{\beta\omega}+1\right)^{-1}$
is the Fermi function, with $\beta$ the inverse temperature, $\mu$
is the chemical potential of the electronic system, 
\begin{equation}
A_{\ell\mathbf{R}_{\ell}\alpha;\ell^{\prime}\mathbf{R}_{\ell^{\prime}}\alpha^{\prime}}(\omega)=\left\langle \ell,\mathbf{R}_{\ell},\alpha\right|\delta\left(\omega-H\right)\left|\ell^{\prime},\mathbf{R}_{\ell^{\prime}},\alpha^{\prime}\right\rangle \label{eq:2pt-spectral_function}
\end{equation}
is the two-point spectral function of crystal bound states in the
Wannier basis and $M_{E,\mathbf{p}\mid\omega_{0},\mathbf{q}\mid\ell\mathbf{R}_{\ell}\alpha}=-2m\left\langle \psi_{E,\mathbf{p}}\right|\mathbf{J}\cdot\mathbf{A}_{\omega_{0},\mathbf{q}}\left|\ell,\mathbf{R}_{\ell},\alpha\right\rangle /\hbar^{2}$
are ARPES matrix elements for the Wannier states, where $\left|\psi_{E,\mathbf{p}}\right\rangle $
is the photoemitted state, $\mathbf{J}$ is the paramagnetic current
operator and $\mathbf{A}_{\omega_{0},\mathbf{q}}$ is the electromagnetic
vector potential. Approximating the photoemitted state by a plane-wave
\cite{Shirley_1995,Amorim_2018} $\psi_{E,\mathbf{p}}(\mathbf{r})\simeq e^{i\mathbf{p}\cdot\mathbf{r}}$
and writing $\mathbf{A}_{\omega_{0},\mathbf{q}}(\mathbf{r})=A_{\omega_{0},\mathbf{q}}^{\lambda}\mathbf{e}_{\mathbf{q},\lambda}e^{i\mathbf{q}\cdot\mathbf{r}}$,
where $\mathbf{e}_{\mathbf{q},\lambda}$ is a polarization vector,
we obtain 
\begin{equation}
M_{E,\mathbf{p}\mid\omega_{0},\mathbf{q}\mid\ell\mathbf{R}_{\ell}\alpha}\simeq\frac{2e}{\hbar}A_{\omega_{0},\mathbf{q}}^{\lambda}\left(\mathbf{p}\cdot\mathbf{e}_{\mathbf{q},\lambda}\right)e^{-i\mathbf{Q}\cdot\left(\mathbf{R}_{\ell}+\bm{\tau}_{\ell,\alpha}\right)}\tilde{w}_{\ell,\alpha}\left(\mathbf{Q}\right),\label{eq:ARPES_matrix}
\end{equation}
where $\mathbf{Q}=\mathbf{p}-\mathbf{q}$ and $\tilde{w}_{\ell,\alpha}\left(\mathbf{Q}\right)$
is the Fourier transform of the Wannier wavefunction of sublattice/orbital
$\alpha$ and layer $\ell$, centred at the origin. In order to evaluate
$A_{\ell\mathbf{R}_{\ell}\alpha;\ell^{\prime}\mathbf{R}_{\ell^{\prime}}\alpha^{\prime}}(\omega)$,
we notice that for each layer, Bloch states form a complete basis,
such that we can write the identity in the space of states including
the three layers as $\text{Id}=\sum_{\ell,\mathbf{k}\in\text{BZ}\ell,\alpha}\left|\ell,\mathbf{k},\alpha\right\rangle \left\langle \ell,\mathbf{k},\alpha\right|$.
Using this fact, we can write
\begin{multline}
A_{\ell\mathbf{R}_{\ell}\alpha;\ell^{\prime}\mathbf{R}_{\ell^{\prime}}\alpha^{\prime}}(\omega)=\\
=\frac{1}{\sqrt{N_{\ell}N_{\ell^{\prime}}}}\sum_{\substack{\mathbf{k}\in\text{BZ}\ell\\
\mathbf{k}^{\prime}\in\text{BZ}\ell^{\prime}
}
}e^{i\mathbf{k}\cdot\left(\mathbf{R}_{\ell}+\bm{\tau}_{\ell,\alpha}\right)}e^{-i\mathbf{k}^{\prime}\cdot\left(\mathbf{R}_{\ell^{\prime}}+\bm{\tau}_{\ell^{\prime},\alpha^{\prime}}\right)}\\
\times\left\langle \ell,\mathbf{k},\alpha\right|\delta\left(\omega-H\right)\left|\ell^{\prime},\mathbf{k}^{\prime},\alpha^{\prime}\right\rangle ,\label{eq:2pt-spectral_function_Bloch}
\end{multline}
where we used the fact that $\left\langle \left.\ell,\mathbf{R}_{\ell},\alpha\right|\ell,\mathbf{k},\alpha\right\rangle =e^{i\mathbf{k}\cdot\left(\mathbf{R}_{\ell}+\bm{\tau}_{\ell,\alpha}\right)}/\sqrt{N_{\ell}}$.
Inserting Eqs.~(\ref{eq:ARPES_matrix}) and (\ref{eq:2pt-spectral_function_Bloch})
into Eq.~(\ref{eq:ARPEs_fermi_golden_rule}), and performing the
sums over the Bravais lattice sites $\mathbf{R}_{\ell}$ and $\mathbf{R}_{\ell^{\prime}}$
the following expression is obtained
\begin{widetext}

\begin{multline}
I_{\text{ARPES}}\left(E,\mathbf{p}|\omega_{0},\mathbf{q}\right)\propto f\left(\omega-\mu\right)\left|\frac{2e}{\hbar}A_{\omega_{0},\mathbf{q}}^{\lambda}\right|^{2}\left|\mathbf{p}\cdot\mathbf{e}_{\mathbf{q},\lambda}\right|^{2}\sum_{\substack{\ell,\mathbf{k}\in\text{BZ}\ell,\alpha,\mathbf{G}_{\ell}\\
\ell^{\prime},\mathbf{k}^{\prime}\in\text{BZ}\ell^{\prime},\alpha^{\prime},\mathbf{G}_{\ell^{\prime}}
}
}\sqrt{N_{\ell}N_{\ell^{\prime}}}\tilde{w}_{\ell,\alpha}\left(\mathbf{Q}\right)\tilde{w}_{\ell^{\prime},\alpha^{\prime}}^{*}\left(\mathbf{Q}\right)\\
\times\delta_{\mathbf{k}-\mathbf{Q}_{\perp},\mathbf{G}_{\ell}}\delta_{\mathbf{k}^{\prime}-\mathbf{Q}_{\perp},\mathbf{G}_{\ell^{\prime}}}e^{-iQ_{z}\tau_{\ell,\alpha}^{z}}e^{i\mathbf{G}_{\ell}\cdot\bm{\tau}_{\ell,\alpha}}\left\langle \ell,\mathbf{k},\alpha\right|\delta\left(\omega-H\right)\left|\ell^{\prime},\mathbf{k}^{\prime},\alpha^{\prime}\right\rangle e^{-i\mathbf{G}_{\ell^{\prime}}\cdot\bm{\tau}_{\ell^{\prime},\alpha^{\prime}}}e^{iQ_{z}\tau_{\ell^{\prime},\alpha^{\prime}}^{z}},
\end{multline}
where $\mathbf{Q}_{\perp}$ is the projection of $\mathbf{Q}$ in
the plane. Using the fact that for a reciprocal lattice vector $\mathbf{G}_{\ell}$
we have $\left|\ell,\mathbf{k},\alpha\right\rangle =e^{i\mathbf{G}_{\ell}\cdot\bm{\tau}_{\ell,\alpha}}\left|\ell,\mathbf{k}-\mathbf{G}_{\ell},\alpha\right\rangle $,
we can use the Kronecker symbols to perform the sum over $\mathbf{k}$,
$\mathbf{G}_{\ell}$ and $\mathbf{k}^{\prime}$, $\mathbf{G}_{\ell^{\prime}}$.
We obtain
\begin{multline}
I_{\text{ARPES}}\left(E,\mathbf{p}|\omega_{0},\mathbf{q}\right)\propto f\left(\omega-\mu\right)\left|\frac{2e}{\hbar}A_{\omega_{0},\mathbf{q}}^{\lambda}\right|^{2}\left|\mathbf{p}\cdot\mathbf{e}_{\mathbf{q},\lambda}\right|^{2}A\sum_{\substack{\ell,\alpha\\
\ell^{\prime},\alpha^{\prime}
}
}\frac{1}{\sqrt{A_{\text{u.c.}\ell}A_{\text{u.c.}\ell^{\prime}}}}e^{-iQ_{z}\tau_{\ell,\alpha}^{z}}\tilde{w}_{\ell,\alpha}\left(\mathbf{Q}\right)e^{iQ_{z}\tau_{\ell^{\prime},\alpha^{\prime}}^{z}}\tilde{w}_{\ell^{\prime},\alpha^{\prime}}^{*}\left(\mathbf{Q}\right)\\
\times\left\langle \ell,\mathbf{Q}_{\perp},\alpha\right|\delta\left(\omega-H\right)\left|\ell^{\prime},\mathbf{Q}_{\perp},\alpha^{\prime}\right\rangle ,
\end{multline}
where we used the fact that $N_{\ell}A_{\text{u.c.}\ell}=N_{\ell^{\prime}}A_{\text{u.c.}\ell^{\prime}}=A$,
the total area of the structure \cite{Koshino_2015}. The remaining
task is to evaluate $\left\langle \ell,\mathbf{Q}_{\perp},\alpha\right|\delta\left(\omega-H\right)\left|\ell^{\prime},\mathbf{Q}_{\perp},\alpha^{\prime}\right\rangle $.
Using the method of the previous section, we construct the matrix
$\bm{H}_{\mathbf{Q}_{\perp}}^{\text{umklapp}}$. Having obtained its
eigenstates and eigenvectors, we can compute
\begin{equation}
\left\langle \ell,\mathbf{Q}_{\perp},\alpha\right|\delta\left(\omega-H\right)\left|\ell^{\prime},\mathbf{Q}_{\perp},\alpha^{\prime}\right\rangle =\sum_{n}\phi_{\ell,\mathbf{Q}_{\perp},\alpha}^{n}\left(\mathbf{0},\mathbf{0}\right)\left[\phi_{\ell^{\prime},\mathbf{Q}_{\perp},\alpha^{\prime}}^{n}\left(\mathbf{0},\mathbf{0}\right)\right]^{*}\delta\left(\omega-E_{\mathbf{k},n}\right),
\end{equation}
which allows us to write 
\begin{equation}
I_{\text{ARPES}}\left(E,\mathbf{p}|\omega_{0},\mathbf{q}\right)\propto f\left(\omega-\mu\right)\left|\frac{2e}{\hbar}A_{\omega_{0},\mathbf{q}}^{\lambda}\right|^{2}\left|\mathbf{p}\cdot\mathbf{e}_{\mathbf{q},\lambda}\right|^{2}A\sum_{n}\left|\mathcal{M}_{\mathbf{Q}_{\perp},n}\right|^{2}\delta\left(\omega-E_{\mathbf{Q}_{\perp},n}\right),
\end{equation}
where 
\begin{equation}
\mathcal{M}_{\mathbf{Q},n}=\sum_{\ell,\alpha}\frac{1}{\sqrt{A_{\ell}}}e^{-iQ_{z}\tau_{\ell,\alpha}^{z}}\tilde{w}_{\ell,\alpha}\left(\mathbf{Q}\right)\phi_{\ell,\mathbf{Q}_{\perp},\alpha}^{n}\left(\mathbf{0},\mathbf{0}\right),
\end{equation}
is the ARPES visibility amplitude for state $\left|\psi_{\mathbf{Q}_{\perp},n}^{\text{umklapp}}\right\rangle $.
As for the bilayer case, the ARPES amplitude only depends on the eigenstate
coefficients $\phi_{\ell,\mathbf{k},\alpha}^{n}\left(\mathbf{0},\mathbf{0}\right)$\cite{Amorim_2018}.
\end{widetext}

\subsubsection{Local density of states }

The local density of states is given by the same site, two-point spectral
function, Eq.~(\ref{eq:2pt-spectral_function}), $\text{LDoS}_{\ell,\mathbf{R}_{\ell},\alpha}(\omega)=A_{\ell\mathbf{R}_{\ell}\alpha;\ell\mathbf{R}_{\ell}\alpha}(\omega)$.
Using the representation of the identity in terms of Bloch states
of individual layers, we can write the local density of states in
the form of Eq.~(\ref{eq:2pt-spectral_function_Bloch}), with $\ell,\mathbf{R}_{\ell},\alpha=\ell^{\prime},\mathbf{R}_{\ell^{\prime}},\alpha^{\prime}$.
The quantity $\left\langle \ell,\mathbf{k},\alpha\right|\delta\left(\omega-H\right)\left|\ell^{\prime},\mathbf{k}^{\prime},\alpha^{\prime}\right\rangle $
can be evaluated by constructing the matrices $\bm{H}_{\mathbf{k}}^{\text{umklapp}}$
and $\bm{H}_{\mathbf{k}^{\prime}}^{\text{umklapp}}$ and obtaining
the corresponding eigenstates and energies, from which we can write
(focusing on layer $1$)
\begin{widetext}
\begin{align}
\left\langle 1,\mathbf{k},\alpha\right|\delta\left(\omega-H\right)\left|1,\mathbf{k}^{\prime},\alpha\right\rangle  & =\frac{1}{2}\sum_{n,\mathbf{G}_{1},\mathbf{G}_{2},\mathbf{G}_{3}}e^{i\mathbf{G}_{1}\cdot\bm{\tau}_{1,\alpha}}\phi_{1,\mathbf{k},\alpha}^{n}\left(\mathbf{0},\mathbf{0}\right)\left[\phi_{1,\mathbf{k},\alpha}^{n}\left(\mathbf{G}_{2},\mathbf{G}_{3}\right)\right]^{*}\delta_{\mathbf{k}^{\prime}-\mathbf{k},\mathbf{G}_{1}+\mathbf{G}_{2}+\mathbf{G}_{3}}\delta\left(\omega-E_{\mathbf{k},n}\right)\nonumber \\
 & +\frac{1}{2}\sum_{n,\mathbf{G}_{1},\mathbf{G}_{2},\mathbf{G}_{3}}e^{-i\mathbf{G}_{1}\cdot\bm{\tau}_{1,\alpha}}\phi_{1,\mathbf{k}^{\prime},\alpha}^{n}\left(\mathbf{G}_{2},\mathbf{G}_{3}\right)\left[\phi_{1,\mathbf{k}^{\prime},\alpha}^{n}\left(\mathbf{0},\mathbf{0}\right)\right]^{*}\delta_{\mathbf{k}-\mathbf{k}^{\prime},\mathbf{G}_{1}+\mathbf{G}_{2}+\mathbf{G}_{3}}\delta\left(\omega-E_{\mathbf{k}^{\prime},n}\right),
\end{align}
where we used that fact that $\left\langle \left.1,\mathbf{k}+\mathbf{G}_{2}+\mathbf{G}_{3},\alpha\right|1,\mathbf{k}^{\prime},\alpha\right\rangle =\sum_{\mathbf{G}_{1}}e^{i\mathbf{G}_{1}\cdot\bm{\tau}_{1,\alpha}}\delta_{\mathbf{k}^{\prime}-\mathbf{k},\mathbf{G}_{1}+\mathbf{G}_{2}+\mathbf{G}_{3}}$.
Inserting this into Eq.~(\ref{eq:2pt-spectral_function}) and using
the Kronecker symbol to perform the sum over $\mathbf{k}$ or $\mathbf{k}^{\prime}$
and $\mathbf{G}_{1}$, we obtain
\begin{equation}
\text{LDoS}_{1,\mathbf{R}_{1},\alpha}(\omega)=A_{\text{u.c.1}}\int_{\text{BZ}1}\frac{d^{2}\mathbf{k}}{\left(2\pi\right)^{2}}\sum_{n}\text{Re}\left\{ \sum_{\mathbf{G}_{2},\mathbf{G}_{3}}e^{i\left(\mathbf{G}_{2}+\mathbf{G}_{3}\right)\cdot\left(\mathbf{R}_{1}+\bm{\tau}_{1,\alpha}\right)}\phi_{1,\mathbf{k},\alpha}^{n}\left(\mathbf{G}_{2},\mathbf{G}_{3}\right)\left[\phi_{1,\mathbf{k},\alpha}^{n}\left(\mathbf{0},\mathbf{0}\right)\right]{}^{*}\right\} \delta\left(\omega-E_{\mathbf{k},n}\right),
\end{equation}
where we transformed the sum over $\mathbf{k}$ into an integral $\sum_{\mathbf{k}\in\text{BZ}\ell}=N_{1}A_{\text{u.c.\ensuremath{\ell}}}\int_{\text{BZ}1}\frac{d^{2}\mathbf{k}}{\left(2\pi\right)^{2}}$.
Similar expressions are obtained for the other layers. Notice that
the local density of states for sites on layer $\ell$ is obtaining
by integrating $\mathbf{k}$ over the Brillouin zone of layer $\ell$. 
\end{widetext}

\begin{figure}
\begin{centering}
\includegraphics[width=8cm]{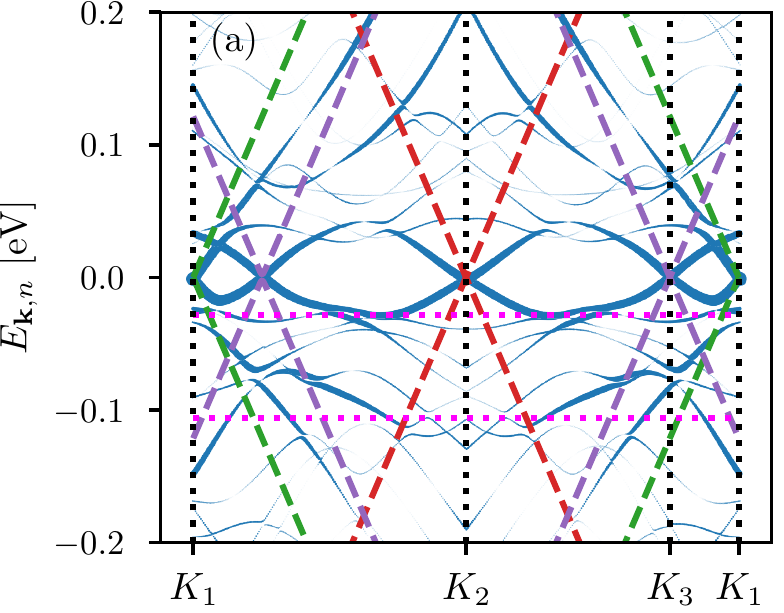}
\par\end{centering}
\begin{centering}
\includegraphics[width=8cm]{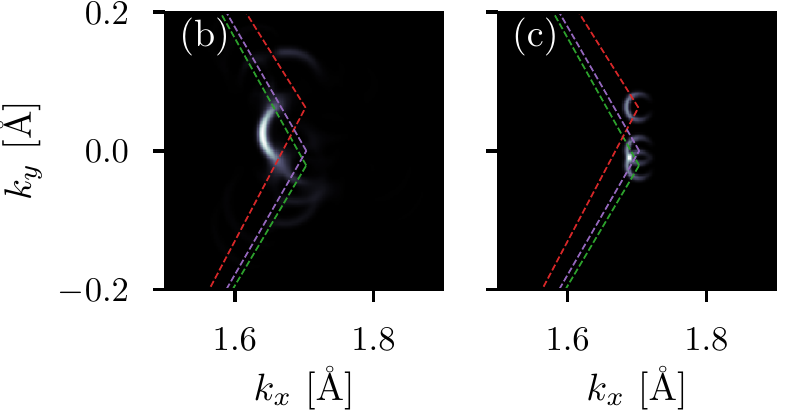}
\par\end{centering}
\caption{\label{fig:ARPES}(a) ARPES mapped band structure of tTLG with $\theta_{1}=-0.71^{\circ}$,
$\theta_{2}=2.1^{\circ}$ and $\theta_{3}=0^{\circ}$. The bands are
shown along the path $\text{K}_{1}\rightarrow\text{K}_{2}\rightarrow\text{K}_{3}\rightarrow\text{K}_{1}$,
where $\text{K}_{\ell}$ is the Dirac point of layer $\ell$. The
thickness of the blue lines is proportional to $\left|\mathcal{M}_{\mathbf{Q}_{\perp},n}\right|^{2}$,
corresponding to the visibility of the bands in ARPES. The dashed
lines represent the band structure of the three decoupled graphene
layers, following the colour code: layer 1 in green, layer 2 in red,
and layer 3 in purple. The horizontal dotted lines mark the energies
$\omega=-0.106$ eV and $\omega=-0.028$ eV, which correspond to two
van Hove singularities highlighted in Fig.~\ref{fig:TDoS}. (b) ARPES
constant energy map for the same tTLG structure at $\omega=-0.106$
eV. The constant energy map for three decoupled graphene layers is
shown in (c) for comparison. The dashed lines represent the Brillouin
zone of each layer, following the same colour code as in (a). A broadening
of $20$ meV was used.}
 
\end{figure}

\subsubsection{Total density of states}

The total density of states normalized by the total number of states
of the trilayer is given by summing over all local density of states
\begin{equation}
\text{TDoS}(\omega)=\frac{1}{\sum_{\ell}N_{\ell}N_{\text{orb}\ell}}\sum_{\ell,\mathbf{R}_{\ell},\alpha}\text{LDoS}_{\ell,\mathbf{R}_{\ell},\alpha}(\omega).\label{eq:tdos_def}
\end{equation}
Noticing that sums of the form $\sum_{\mathbf{R}_{1}}e^{i\left(\mathbf{G}_{2}+\mathbf{G}_{3}\right)\cdot\mathbf{R}_{1}}=N_{1}\sum_{\mathbf{G}_{1}}\delta_{\mathbf{G}_{1}+\mathbf{G}_{2}+\mathbf{G}_{3},\mathbf{0}}=N_{1}\delta_{\mathbf{G}_{3},\mathbf{0}}\delta_{\mathbf{G}_{2},\mathbf{0}}$,
since for fully incommensurate structures $\mathbf{G}_{1}+\mathbf{G}_{2}+\mathbf{G}_{3}=0$
is only possible if $\mathbf{G}_{1}=\mathbf{G}_{2}=\mathbf{G}_{3}=\mathbf{0}$,
we can perform the sums over $\mathbf{R}_{\ell}$'s in Eq.~(\ref{eq:tdos_def})
obtaining
\begin{multline}
\text{TDoS}(\omega)=\frac{1}{\sum_{\ell}A_{\text{u.c.}\ell}^{-1}N_{\text{orb}\ell}}\\
\times\sum_{\ell}\int_{\text{BZ}\ell}\frac{d^{2}\mathbf{k}}{\left(2\pi\right)^{2}}\sum_{n,\alpha}\left|\phi_{\ell,\mathbf{k},\alpha}^{n}\left(\mathbf{0},\mathbf{0}\right)\right|^{2}\delta\left(\omega-E_{\mathbf{k},n}\right),
\end{multline}
where we used the fact that $N_{\ell}/A=A_{\text{u.c.}\ell}^{-1}$.
The contribution from each layer to the total density of states is
expressed in terms of an integration over the Brillouin zone of that
layer. In the case of a bilayer, the previous result reduces to the
one derived in a mathematically rigorous way in Ref.~\cite{Massatt_2018}. 

\begin{figure}
\begin{centering}
\includegraphics[width=8cm]{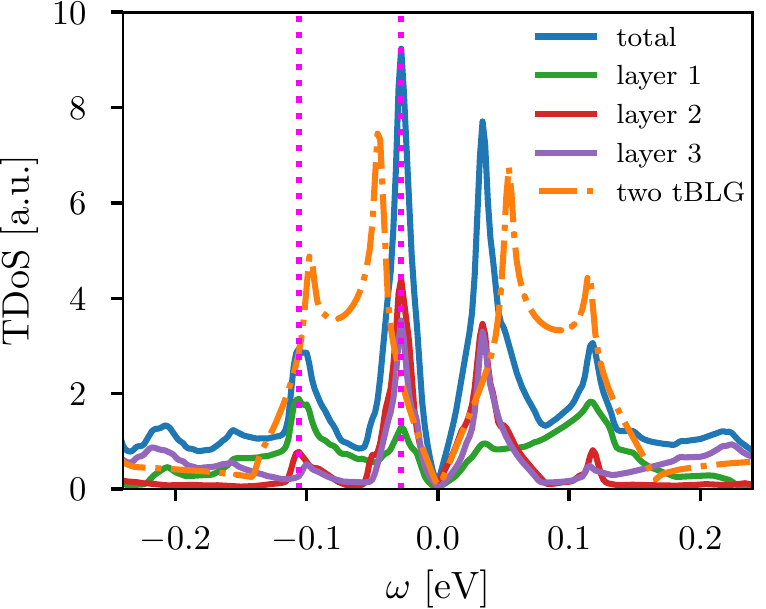}
\par\end{centering}
\caption{\label{fig:TDoS}Total and layer resolved density of states for tTLG,
with $\theta_{1}=-0.71^{\circ}$, $\theta_{2}=2.1^{\circ}$ and $\theta_{3}=0^{\circ}$.
The dot-dashed line shows the total density of states obtained by
modelling the tTLG as two tBLG (with the contribution of layer 2 averaged).
The two dotted vertical lines mark two van Hove singularities at $\omega=-0.106$
eV and $\omega=-0.028$ eV. The calculation was performed using a
mesh of 56677 k points in a circular region of radius $0.043\text{ Å}^{-1}$around
the Dirac points of each layer. A broadening of $2$ meV was used. }
\end{figure}

\section{Application to twisted trilayer graphene\label{sec:Application_tTLG}}

\begin{figure}
\begin{centering}
\includegraphics[width=8cm]{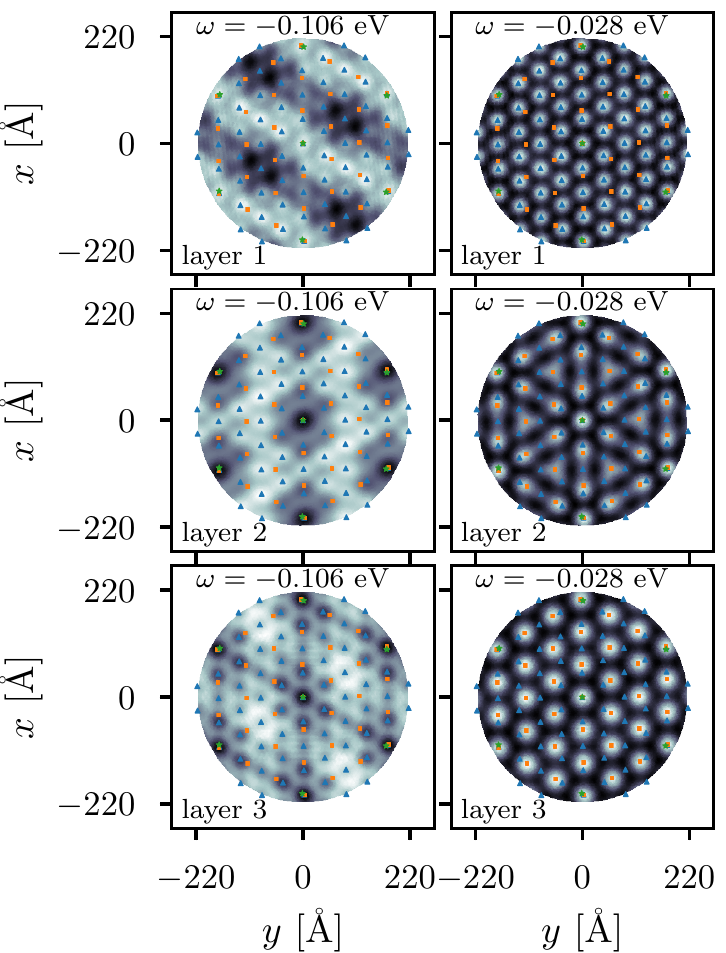}
\par\end{centering}
\caption{\label{fig:LDoS}Layer resolved local density of states for tTLG,
with $\theta_{1}=-0.71^{\circ}$, $\theta_{2}=2.1^{\circ}$ and $\theta_{3}=0^{\circ}$,
at the two van Hove singularities marked by the vertical lines in
Fig.~\ref{fig:LDoS}. Brighter regions correspond to regions with
higher density of states. The blue triangles, yellow squares and green
stars show, respectivelly, the moiré lattices due to the interference
of layers 1 with 2, layers 2 with 3, and layers 1 with 3. The calculation
was performed using a mesh of 439 k points in a circular region of
radius $0.019\text{ Å}^{-1}$around the Dirac points of each layer.
A broadening of $20$ meV was used. }
\end{figure}

We now apply the general formalism developed in the previous section
to the case of incommensurate tTLG. We model individual layers within
the $p_{z}$ orbital, nearest neighbour tight-binding Hamiltonian,
with hopping $-t$. For the interlayer coupling we use a Slatter-Koster
approximation
\begin{equation}
h_{\alpha\beta}^{\ell\ell^{\prime}}\left(\mathbf{R}_{\ell},\mathbf{R}_{\ell^{\prime}}\right)=V_{pp\pi}\left(R\right)\frac{r^{2}}{R^{2}}+V_{pp\sigma}\left(R\right)\frac{d^{2}}{R^{2}},
\end{equation}
where $R=\sqrt{r^{2}+d^{2}}$ is distance between the Wannier centres,
with $r=\left|\mathbf{R}_{\ell}+\bm{\tau}_{\ell,\alpha}^{\perp}-\mathbf{R}_{\ell^{\prime}}-\bm{\tau}_{\ell^{\prime},\beta}^{\perp}\right|$
the in-plane distance and $d=3.35\,\text{Å}$ the interlayer separation.
The Slatter-Koster functions are parametrized as $V_{pp\pi}\left(R\right)=-te^{-\left(R-a_{\text{CC}}\right)/r_{0}}$
and $V_{pp\sigma}\left(R\right)=t_{\perp}e^{-\left(R-d\right)/r_{0}}$,
with $t=2.7$ eV, $t_{\perp}=0.48$ eV, $r_{0}=0.453\,\text{Å}$,
and $a_{\text{CC}}=1.42\,\text{Å}$ the intralayer nearest-neighbour
distance\cite{Moon_2013}. Motivated by the recent experimental work
of Ref.~\cite{Zuo_2018}, we will focus on a tTLG, where the top
layer (layer 1) is rotated by an angle $\theta_{1}=-0.71^{\circ}$,
the middle layer (layer 2) is rotated by an angle $\theta_{2}=2.1^{\circ}$
and the bottom layer (layer 3) is taken as the reference, with $\theta_{3}=0^{\circ}$.
When constructing the Hamiltonian matrix $\bm{H}_{\mathbf{k}}^{\text{umklapp}}$,
we chose a momentum cutoff $\Lambda=2.1\left|\text{K}\right|$, where
$\left|\text{K}\right|=4\pi/\left(3\sqrt{3}a_{\text{CC}}\right)$
is the distance of the Dirac points from the origin, such that the
first star of reciprocal lattice vectors of each layer is included.
In Fig.~\ref{fig:ARPES}, we shown the computed ARPES mapped band
structure and constant energy contour. It is clear that the interlayer
coupling leads to a significant reconstruction of the band structure.
This is further confirmed if we look at the low energy total density
of states, which is shown in Fig.~\ref{fig:TDoS}. As can be seen
the hybridization of layers 1 and 2, and layers 2 and 3 gives origin
to two sets of low energy van Hove singularities. However, and differently
from what is claimed in Ref.~\cite{Zuo_2018}, the trilayer structure
cannot simply be described as two tBLG structures. To show this, in
Fig.~\ref{fig:TDoS} we also present the total density of states
computed by describing the trilayer as two bilayers, with the contribution
form layer 2 averaged between the two bilayer systems. As can be seen,
there is a significant spectral reconstruction in the trilayer. The
presence of the three layers leads to a increased separation between
the van Hove singularities of the two bilayer structures. The importance
of considering the three layers of the tTLG is also shown when studying
the local density of states of the system, which we show in Fig.~\ref{fig:LDoS},
at the energies corresponding to the van Hove singularities marked
in \ref{fig:TDoS}. It is clear that the layer resolved LDoS displays
a modulation corresponding to the expected moiré pattern due to interference
of layer 1 with 3, and layer 2 with 3. However, an additional modulation
is observed that corresponds to a moiré pattern due to the interference
between layers 1 with 3. This is specially clear in the LDoS of layer
3 at $\omega=-0.106$ eV, which displays a clear modulation with the
periodicity of the moiré lattice due to the interference of layers
1 and 3 (whose corresponding lattice is represented by the green star
markers). This effect can only be captured if considering coupling
between the three layers simultaneously. 

\section{Conclusions\label{sec:Conclusions}}

In this work, we have developed a tight-binding based, momentum space
formalism to describe the electronic properties of incommensurate
multilayer van der Waals structures. The method is based on an expansion
of the electronic wavefunction in terms of Bloch waves of individual
layers, including generalized umklapp scattering due to the competition
between the periodicities of the different layers. We also showed
how the momentum resolved, local and total density of states, which
can be measured via ARPES and STS, can be computed using the developed
formalism. Interestingly, both the total and the local density of
states can be expressed in terms of integrals over the Brillouin zone
of the different layers, a result previously obtained for the total
density of states in the bilayer case \cite{Massatt_2018}. We applied
the general formalism to study the spectral properties of tTLG. We
found out that the coupling between the three layers can significantly
affect the low energy spectral properties, which cannot be simply
attributed to the pairwise hybridization between the layers. We found
that the low energy van Hove singularities due to the coupling between
consecutive layers are repelled due to the hybridization between the
three layers. This hybridization between the three layers is also
manifested in the modulation of the LDoS, which, besides the moiré
patterns due to layers 1 with 2, and layers 2 with 3, also display
a modulation due to the hybridization between layers 1 with 3. The
formalism developed in this paper is capable of describing structures
with arbitrary lattice mismatch and misalignment. Its flexibility
makes it very promising to study spectral and transport properties
of the technologically relevant graphene/boron nitride/graphene and
graphene/STMD/graphene structures.
\begin{acknowledgments}
B. A. received funding from the European Union\textquoteright s Horizon
2020 research and innovation programme under grant agreement No 706538.
E. V. C. acknowledges partial support from FCT-Portugal through Grant
No. UID/CTM/04540/2013.
\end{acknowledgments}

%

\end{document}